\documentclass[
fleqn,usenatbib]{mnras}

\usepackage{newtxtext,newtxmath}

\usepackage[T1]{fontenc}
\usepackage{ae,aecompl}

\usepackage{graphicx}   
\usepackage{amsmath}    
\usepackage{amssymb}    

\title[Cyclic variations of solar magnetic field]{Cyclic variations in the main components of the solar large-scale magnetic field}

\author[V. N. Obridko et al.]{
V.N.Obridko,$^{1}$\thanks{E-mail: obridko@mail.ru}
D.D.Sokoloff,$^{1,2}$
B.D.Shelting,$^{1}$
A.S.Shibalova,$^{1,2}$
and I.M.Livshits$^{1,3}$
\\
$^{1}$IZMIRAN, Kaluzhskoe shosse, 4, Troitsk, Moscow,  142191, Russia\\
$^{2}$Department of Physics, Moscow State University, Moscow, 119991, Russia\\
$^{3}$ Sternberg State Astronomical Institute, Lomonosov Moscow State University, Moscow, 119991, Russia
}

\date{Accepted 2019 January 14. Received 2020 January 14; in original form 2019 October 21}

\pubyear{2019}

\begin{document}
\label{firstpage}
\pagerange{\pageref{firstpage}--\pageref{lastpage}}

\maketitle

\begin{abstract}
We considered variations the dipole and the quadrupole components of
the solar large-scale magnetic field. Both the axial and the
equatorial dipoles exhibit a systematic decrease in the past four
cycles in accordance with the general decrease of solar activity.
The transition of the pole of a dipole from the polar region to mid
latitudes occurs rather quickly, so that the longitude of the pole
changes little. With time, however, this inclined dipole region
shifts to larger longitudes, which suggests an acceleration of the
dipole rotation. The mean rotation rate exceeds the Carrington
velocity by 0.6\%. The behavior of the quadrupole differs
dramatically. Its decrease over last four cycles was much smaller
than that of the dipole moment. The ratio of the quadrupole and
dipole moments has increased for four cycles more than twice in
contrast to the sunspot numbers, which displayed a twofold decrease
for the same time interval.
What about quadrupole rotation, the
mean longitude of  the poles of one sign decreased by 600
degrees over four cycles, which suggests that the mean rotation rate
was lower than the Carrington velocity by 0.28\%.
We
do not see  however any conclusive evidence that, in the period under
discussion, a mode of quadrupole symmetry was excited on the Sun
along with the dipole mode.

\end{abstract}




\section{Introduction}

It is generally agreed that most of the principal active phenomena in the Sun are due to the evolution of local magnetic fields. This evolution results in formation of sunspots and active regions, i.e., the features of a few arc. sec. in size. The solar features of larger scale are not seen on high-resolution images; however, they are clearly revealed on Stanford synoptic charts, whose
resolution is $3'\times 3'$. On magnetic charts, one can conventionally isolate local (a few arc. min.) and large-scale (from several  to $10' \times 10'$ to  $20' \times 20'$, i.e., $\approx 0.2 \, R_\odot$) magnetic fields. A
large-scale field involves two high-latitude regions occupied by the fields of opposite sign and a set of quasi-unipolar structures at the mid and low latitudes. Quasi-unipolar magnetic regions, which
are sometimes the remnants of large active regions, live from half a year to 1-2 years. The largest scale is associated with the dipole component of the magnetic field - the global dipole.

It has long been known, above all, from eclipse observations that at very low activity, the magnetic field in the corona is, under certain approximation, the global dipole, whose axis virtually coincides
with the rotation axis of the Sun. Here, we mean the coronal layers where the solar wind speed is still small, i.e., the layers below the solar wind "source surface". In this phase of the cycle, the disturbing
effect of local fields on the dipole is weak. Such a magnetic configuration remains stable for a few years. Due to the dipole configuration of the global field, a number of high coronal loops connecting the
fields of opposite polarity on both sides of the equator arise in the low-latitude zone. This forms the well-known large-scale structure of the solar corona at the minimum, which includes a belt of streamers
at all longitudes in the vicinity of the equator and two systems of the polar plumes. The global magnetic fields carried out to interplanetary space by the solar wind form the heliospheric current sheet,
which separates the oppositely directed magnetic fluxes. In the periods of low activity, this neutral current sheet is flat. With the rise of activity, the influence of active regions increases, and the
polarity separation line between the large-scale fields in the vicinity of the equator becomes wavy. In addition, the role of the quadrupole and higher-order harmonics of large-scale magnetic fields also
increases in this period.

It is convenient to estimate the contribution of magnetic fields of different scales by studying cyclic variations in the characteristics of different multipoles. In particular, it is important that their
rotation be analysed separately. The dependence of the dynamo characteristics on rotation was considered in Noyes et al. (1984), Saar and Brandenburg (1999), Pizzolato et al. (2003), B\"ohm-Vitense (2007),
Reiners et al. (20090. Some authors have shown that there is also an inverse effect -- negative correlation between the magnetic field and rotation velocity (Hathaway \& Wilson, 1990, Kambry \&
Nishikawa, 1990, Howard, 1984,  Gigolashvili \& Khutsishvili, 1990, Obridko \& Shelting, 2016).

The relationship between the fields of different scales and the characteristics of solar activity was studied by considering mainly the low-order multipoles -- the dipole and quadrupole. Stenflo \&
Vogel (1986), Stenflo \& Weisenhorn (1987), Stenflo \& G\"udel (1988),  and Knaack \& Stenflo (2005) arrived at the conclusion that large-scale fields are based on spherical antisymmetric
harmonics such as the axial dipole and octupole.

Livshits \& Obridko (2006) used data on large-scale magnetic fields (synoptic charts) and the field of the Sun as a star (general magnetic field) to determine the magnetic moment and direction of the
dipole field for the past three solar cycles (21-23). It was found that both the magnitude of the moment and its vertical and horizontal components were changing regularly during a cycle never turning to zero.
The process of polarity reversal of the global dipole is a change in the inclination of its axis, which is not smooth, but occurs in a few stages lasting 1-2 years. Before the reversal begins, the
dipole axis executes a precession about the solar rotation axis and, then, moves in the meridional plane to reach very low latitudes, where it begins to shift significantly in longitude. The vector of
the global dipole precesses (not too regularly) about the direction of the rotation axis. Then, during the polarity reversal, it turns round virtually in the meridian plane, i.e., at the same
heliolongitude in each hemisphere. The amplitudes of the two components change differently. The vertical component changes smoothly reaching the maximum absolute value at the minimum of the cycle
(vertical dipole). The horizontal component determines the maximum value of the total dipole field in the epochs of high activity. It is more variable and actually determines the direction
of the full vector of the dipole field in the process of the polarity reversal.

The characteristics both of the full vector of the dipole moment and of its components change depending on the phase of the cycle.

In addition to the main oscillation mode -- quasi-eleven-year cycle -- there, apparently, exist shorter oscillation periods. Fluctuations of the magnetic moment of the horizontal and vertical
dipoles during a quasi-eleven-year cycle are exactly the same. Livshits \& Obridko (2006) believe that a complete coincidence both in period and in amplitude indicates to essentially the
same physical phenomenon, and the separation into two types does not have much physical sense from the point of view of the main cycle of solar activity. So, we are dealing here with cyclic
changes in the latitude of the inclined rotator. The situation is completely different when fluctuations with periods of 1.3 -- 2.5 years are concerned. These modes are present only in oscillations
of the magnetic moment of the horizontal dipole and are absent in oscillations of the vertical dipole. Their existence is responsible for variations in the tilt of the total dipole.

Similar results were reported by DeRosa et al. (2012). These authors used the same observation data
from the John Wilcox Solar Observatory (WSO) as Livshits \& Obridko
(2006) and coefficients obtained  by using the Potential-Field
Source-Surface Model (PFSS). The data covered three activity cycles
(21, 22, and 23), and the problem was considered much broadly,
including higher-order harmonics.

The evolution of the first harmonics of the solar magnetic field is not only interesting in itself, but is also of particular value as a basis for comparing theoretical ideas in the study of the solar
dynamo with observational data. The fact is that the solar dynamo is operating somewhere under the solar surface. It is generally believed that this occurs in the inversion layer (e.g.,
Choudhuri, 1990); however, it may also be a layer not very far from the surface (e.g. Brandenburg, 2005). It is possible that there are two zones of generation. However, the magnetic fields
that are seen on the surface in the form of sunspots and the large-scale magnetic field differ significantly in their configuration from those that are directly generated by the solar dynamo.
For example, an isolated sunspot is a local feature and does not belong to the surface large-scale magnetic field, while all sunspots together for a period comparable with an 11-year cycle form (as seen on butterfly diagrams) a large-scale structure, which gives us an idea of evolution of the toroidal magnetic field. Therefore, a comparison with the solar dynamo models requires that we considered the evolution of the mean parameters of the solar surface magnetic field.

The basic concepts of the solar dynamo theory were formulated in terms of the mean magnetic field (e.g., Krause \& R\"adler, 1980). At present, of course, a detailed numerical simulation
of the solar dynamo is possible (e.g., Brandenburg et al., 2012); however, when interpreting the results of such a simulation, we have to turn to the behavior of the mean large-scale field parameters.
In this context it is necessary to distinguish between the mean and the instantaneous values of the solar magnetic dipole moment during its inversion (Moss et al., 2013; Pipin et al., 2014), which
emphasizes the complexity of the problem under consideration.

Of particular interest to the dynamo theory is the behavior of the two first harmonics (odd and even) of the solar magnetic field, i.e., the dipole and the quadrupole. The fact is that there are two
types of magnetic fields (odd and even with respect to the solar equator) that can be excited in the convection shell strictly symmetric about the equator. In the process, neither the meridional
circulation, nor the differential rotation and the mirror asymmetry of motions mix these fields. Of course, the odd and even configurations cannot be reduced to a pure dipole or quadrupole if only
because they include a toroidal magnetic field. However, the behavior of the magnetic dipole and quadrupole can illustrate the evolution of these configurations in time.

In particular, in the simplest models of the solar dynamo, both the quadrupole moment and the higher-order even modes are exactly zero. A minor change in the hydrodynamics of the spherical
shell will allow us to simulate generation of magnetic fields of dipolar rather than quadrupolar symmetry (see (Moss et al., 2008) and references therein). It would be natural to suggest that
there may be stars, in which this type of the dynamo mechanism operates.  Possibly, recent Zeeman-Doppler observations of complex structures of stellar magnetic fields  \cite{Retal16}, \cite{Retal18} support this expectation however this point deserves to be addressed in a special paper.
In principle, the dynamo mechanism can
also produce mixed parity solutions (e.g., Jennings\& Weiss, 1991). This is, probably, how the solar dynamo worked at the end of the Maunder minimum (see Sokoloff \& Nesme-Ribes, 1994, Usoskin et al., 2015) and references therein).

The magnetic field of the Sun, obviously, contains the modes both of dipolar, and of quadrupolar symmetry. In principle, we could suggest that the solar dynamo constantly produces a mixed parity
configuration, but at the end of the Maunder minimum, the contribution of quadrupolar modes was abnormally large. However, another point of view is possible, according to which the quadrupolar
modes arise simply because the Sun is not perfectly symmetric with respect to the solar equator. Checking which hypothesis is better supported by observations will, obviously, require a study of the time evolution of the solar quadrupolar moment.

In this paper, which is, in some sense, a development of the ideas of Livshits \& Obridko (2006), we will focus on the analysis of the joint evolution of the basic odd (dipole) and basic even
(quadrupole) harmonics. It is important to note that, now, we have at our disposal data for four activity cycles.

\section{Formulation of the problem and basic equations}

Using WSO synoptic charts of the radial component of the solar magnetic field (Scherrer et al., 1977), we calculated the magnetic field in the potential approximation by the well-known method described in (Hoeksema \& Scherrer, 1986; Hoeksema, 1991) in its classical version without assuming radial field in the photosphere. WSO measurements of the magnetic-field longitudinal component (\url{http://wso.stanford.edu/synopticl.html}) were used as the source data to plot the synoptic charts for each Carrington rotation. The WSO data used in this study cover the time interval of 43 years from the beginning of Carrington rotation (CR) 1642 (27 May 1976) to the end of CR 2210 (23 November 2018).

The coronal magnetic field was extrapolated by solving the boundary problem with the line-of-sight field component measured in the photosphere and a strictly radial field at the source surface. As the source surface we have taken a conventional boundary, where the potential approximation ceases to be true and the field lines are carried away radially by the solar wind. At this boundary, the magnetic field is assumed to be normal to the surface, and the potential is zero. Of course, the position of this conventional boundary can be fixed only approximately and is usually assumed to be located at a distance of 2.5 radii from the center of the Sun. This allowed us to calculate three magnetic-field components in the spherical coordinates $B_r$, $B_\theta$, and $B_\varphi$. The magnetic-field components have the form:

\begin{eqnarray}
B_r=\sum P_n^m (\cos \theta) (g_{nm} \cos m\varphi +h_{nm} \sin m\varphi) \times\\
\times ((n+1)(R_\odot/R)^{n+2}-n(R/R_s)^{n-1}c_n ), \nonumber
\end{eqnarray}

\begin{eqnarray}
B_\theta = - \sum {{\partial P_n^m (\cos \theta)} \over {\partial \theta}} (g_{nm} \cos m\varphi + h_{nm} \sin m \varphi) \times \\
\times ((R_\odot /R)^{n+2} + (R /R_s)^{n-1} c_n), \nonumber
\end{eqnarray}

\begin{eqnarray}
B_\varphi = - \sum {m \over {\sin \theta}} P_n^m (\cos \theta) (h_{nm} \cos m \varphi - g_{nm} \sin m \varphi) \times \\
\times ((R_\odot /R)^{n+2} + (R/R_s)^{n-1} c_n). \nonumber
\end{eqnarray}
Here, $0 \le m \le n \le N$ (usually, $N=9$ ); $c_n= -(R_\odot /R_s)^{n+2}$, where $R_\odot$ and $R_s$  are the solar radius and the radius of the source surface, respectively, measured from the center of the Sun;
$P_n^m$ are the Legendre polynomials; and $g_{nm}$  and $h_{nm}$ are the coefficients of the spherical harmonic analysis obtained by comparison with observations at the photospheric level. It is usually assumed that
$R_\odot=1$ and $R_s=2.5$. It is important to note that the coefficients were calculated under the assumption that the field is potential throughout the photosphere up to the source surface, including the boundaries. At the source surface, the field is assumed strictly radial. The source surface is taken at a distance of 2.5 solar radii from the center of the Sun. The results obtained according to this scheme on the site \url{http://wso.stanford.edu/synopticl.html} are called "classical".

In fact, our calculations yield the corresponding harmonics rather than the classical multipoles. As is known, the direct WSO measurements provide the longitudinal component of the magnetic field. In order to obtain the
expansion coefficients, we find the expression for the longitudinal component from equations (1-3). In this expression, the coefficients are not yet been determined. Then, the expansion coefficients are found by assuming
$R=R_\odot$ and taking into account the orthogonality of the polynomials or directly by applying the least square method. The coefficients obtained, naturally, differ from the coefficients of the contribution of the
classical multipoles. One cannot expect that such a distorted multipole would rotate as a single piece, since the terms determining the dependence on the source surface radius are involved in calculations of the
coefficients. They are different in Eq.\ (1) and Eqs.\ (2-3). The ratio of the components varies with height; therefore, the rotation of the "dipole" cannot be identified with the rotation of the pole. However, these
differences are small enough, and we will continue using the expressions "dipole" and "quadrupole" instead of the more precise terms "first harmonic" and "second harmonic".

Some authors (see the discussion in Wang \& Sheeley (1992)) have pointed out the shortcomings of the classical method and proposed the hypothesis of radial magnetic field in the photosphere. The "radial" computation assumes that the field in the photosphere is radial. Our estimates Obridko et al. (2006) have shown that when these two methods are used, differences do exist and mainly concern the magnetic field intensity. As a rule, the first harmonics are much larger than in the classical method. On the other hand, the differences in the structure of the field lines are insignificant, especially over long time intervals. It can therefore be assumed that the rotation characteristics we find do not strongly depend on the method applied. A slight difference between the results obtained by these two methods is noticeable at latitudes higher
than $70^\circ$.

\begin{figure}
    \includegraphics[width=\columnwidth]{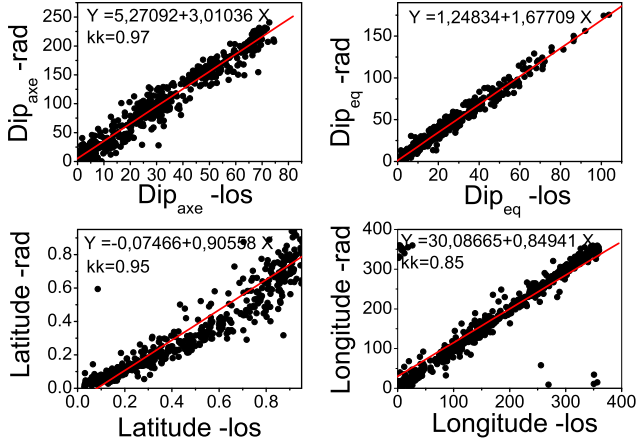}
    \caption{Comparison of the dipole parameters calculated in the classical and radial systems. The field on the top panel is in Gauss.}
    \label{fig1}
\end{figure}

Fig.~1 compares the magnetic moments of the axial (a) and equatorial (b) dipoles, the cosine of latitude (c), and the longitude (d) of the north pole for both systems. The abscissa shows the values in the classical system
(the photospheric surface field is potential; the directly observed longitudinal field is used). The ordinates indicate similar values obtained under the assumption that the field in the photosphere is radial. The
calculations were performed for rotations from 1642 to 2210 without averaging. In all cases, one can see a very high correlation. However, the moment of the axial dipole is three times greater in the case of the radial
hypothesis. The equatorial dipole in the radial hypothesis is greater by a factor of 1.67. The longitudes and latitudes agree. A deviation from the linear dependence for the latitude is due to the fact that the radial
hypothesis underestimates the equatorial dipole.

A new software tool has been developed, which allows
calculating the structure of each harmonic separately. Generally
speaking, the map of a quadrupole sometimes looks rather intricate.
The simplest map is illustrated in Fig.~2.~Two versions of
representation are possible: we can plot either a 3D projection or a
2D map in Carrington coordinates. Fig. 2 shows the maps of the
dipole (2a) and quadrupole (2b,c) for 5 February 2006. On the 3D
maps, the axis is inclined in such a way as an observer might see it
from a point located at $30^\circ$ above the plane of the equator.
On the 2D map, the dotted line shows the cross section of the disk
by the ecliptic plane. The positive polarity (N) is colored blue and
the negative polarity (S), red. The contour lines correspond to the
values of 0, 5, 10, 15, 20, 30, 40, 50, 70, and 100 microtesla.

\begin{figure}
    \Large
    \includegraphics[width=0.48\columnwidth]{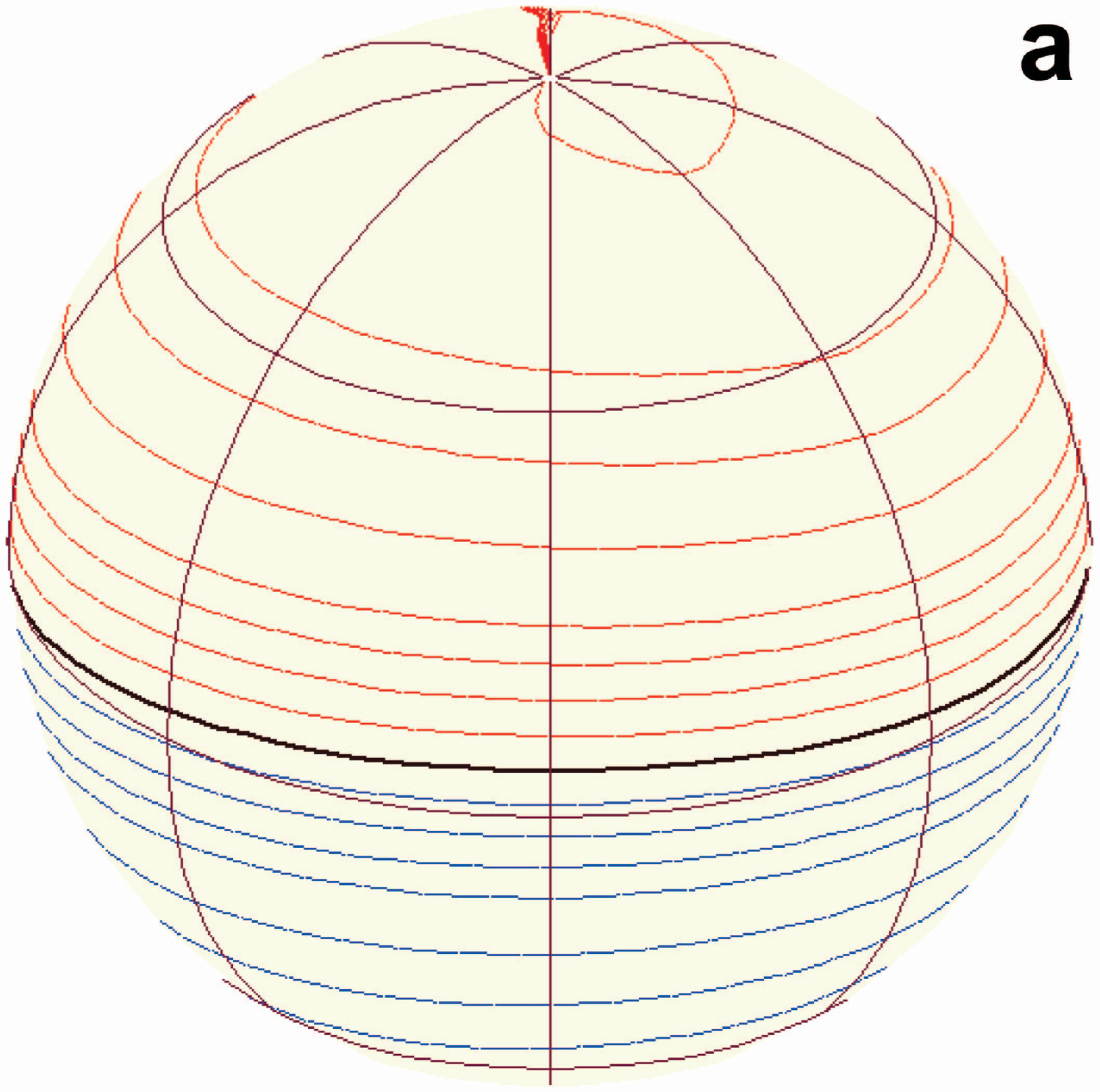}
    \includegraphics[width=0.48\columnwidth]{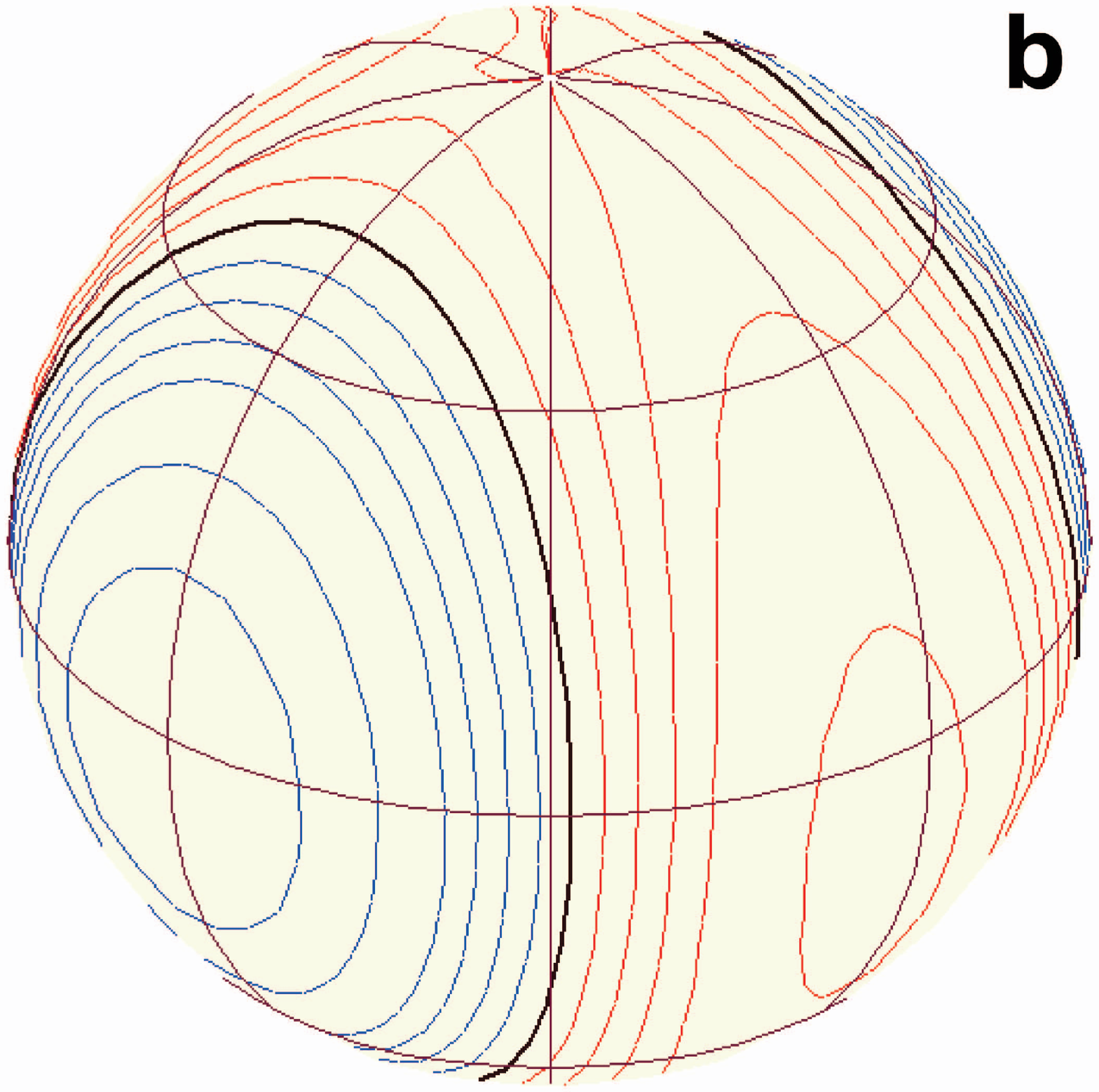}
    \includegraphics[width=\columnwidth]{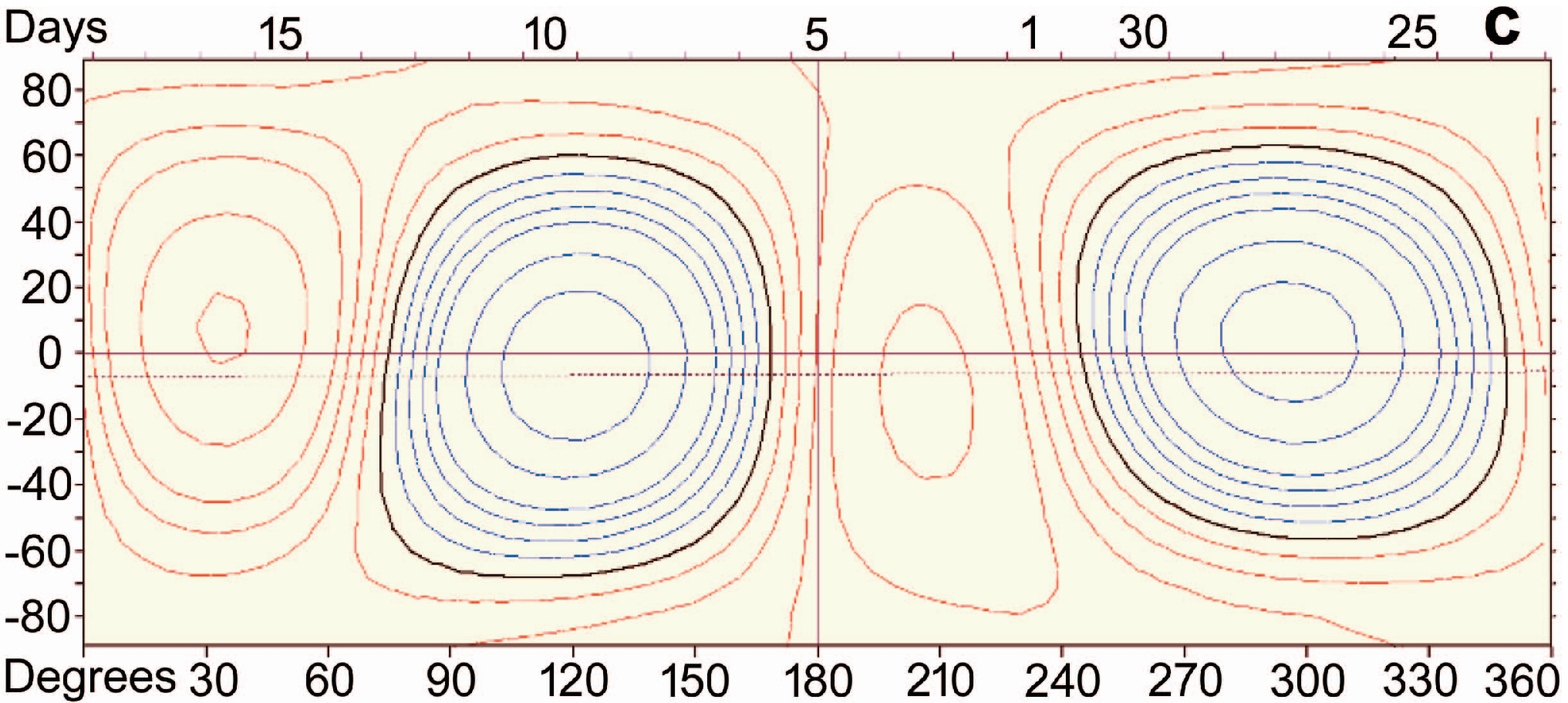}
    \caption{Contour maps of a dipole (a) and quadrupole (b, c). }
    \label{fig2}
\end{figure}

\section{Cyclic variations of a dipole}

In recent years, the dynamo theory and helioseismic studies have made a significant progress. Unfortunately, these studies do not fully take into account the results of observation of magnetic fields in the Sun.
In particular, when analyzing the generation of the magnetic field in the solar interior, it was sometimes assumed that the dipole moment of the Sun disappears completely at the beginning of the field reversal
near the solar maximum and, then, reappears with the opposite sign (Livshits \& Obridko, 2006). This urged us to return to the analysis of contribution of the dipole component to the observed
magnetic field of the Sun (see also Pipin et al., 2014).

The axial dipole is the largest at the minimum of the cycle at high
latitudes. The equatorial dipole reaches its maximum at the maximum
of the cycle, mainly at low latitudes. The cyclic variation in the
values of unsigned magnetic field of the axial, $B_d(ax)$, and
equatorial, $B_d(eq)$, dipoles averaged all over the solar surface
is shown in Fig. 3.

\begin{figure}
    \includegraphics[width=\columnwidth]{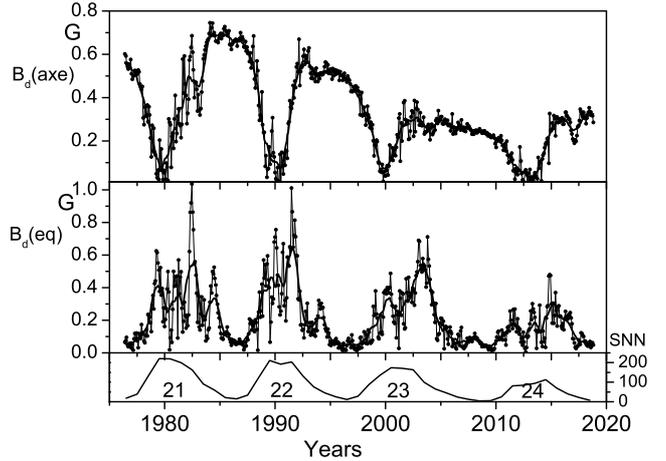}
    \caption{Cyclic variation of the axial, $B_d(ax)$, and equatorial, $B_d(eq)$, dipoles. The lower panel shows the sunspot numbers, SSN (Version 2). }
    \label{fig3}
\end{figure}

As seen from Fig. 3, this simplified description needs
clarification. The three main characteristic points of the cycle do
not coincide. The sunspot maximum of Cycle 21 was recorded in
December 1979. The double maximum of the equatorial dipole (May 1979
and August 1982) does not coincide with this date. Between both
dates, there is a deep gap of the type of the Gnevyshev gap (April
1979). The date of reversal of the axial dipole is close to this gap
(December 1979) and to the date of the sunspot maximum, but is far
ahead of the date of the main maximum of the equatorial dipole. In
Cycle 22, the maximum sunspot numbers (SSN) were recorded in July
1989, which does not coincide with the reversal date (April 1900)
nor with the main (secondary) maximum (July 1991). In Cycle 23, the
SSN maximum took place in April 2000. The reversal almost coincided
with this date, but the maximum of the equatorial dipole was
recorded much later (in April 2003). In Cycle 24, the secondary SSN
maximum was recorded in April 2014, the reversal occurred in March
2013, and the maximum of the axial dipole took place in January
2016. It should also be noted that the reversal of the polar field
lasts rather long, while the sign reversal and, correspondingly, the
minimum in Fig. 3a take relatively short time.

Fig. 4 (upper panel) illustrates variations in the total magnetic
moment of the dipole. The data were smoothed over 14 rotations.

The dipole magnetic moment $M_{\rm dip}$, the axial dipole moment
$M_{\rm ax}$, and the equatorial dipole moment $M_{\rm eq}$ are determined
by the following equations

\begin{eqnarray}
M_{\rm dip}=\sqrt{g^2_{10}+g^2_{11}+h^2_{11}},\\
M_{\rm ax}=|g_{10}|,\\
M_{\rm eq}=\sqrt{g^2_{11}+h^2_{11}}.
\end{eqnarray}

\begin{figure}
    \includegraphics[width=\columnwidth]{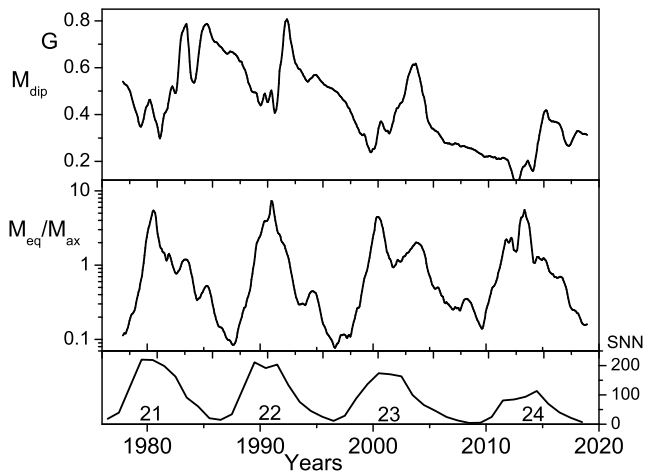}
    \caption{The dipole magnetic moment (upper panel) and the ratio of the equatorial and axial dipole moments (mid panel). The ordinate scale on the mid panel is logarithmic.
     The lower panel shows sunspot numbers (Version 2).}
    \label{fig4}
\end{figure}

 Fig. 4 (mid panel) shows the ratio of the smoothed values of the equatorial and axial dipole moments.
One can see that, in the vicinity of the cycle maximum, the moment
of the equatorial dipole exceeds the moment of the axial dipole by
an order of magnitude, while near the minimum, the ratio is
opposite. Since the dates of the extreme values of the dipole
moments do not coincide, the total dipole not only never drops to
zero, but is significant in magnitude even near the minimum of the
cycle. This led Livshits \& Obridko (2006) to the conclusion that
the dipole turns over and, thus, to the idea of inclined rotator.
This conclusion is true in principle, but not entirely accurate. The
equatorial dipole cannot be considered a direct transform of the
axial dipole. They differ significantly not only in the spatial
orientation, but also in many other properties. In particular, it is
to be noted that the widely known gradual decrease in solar activity
both in the sunspot numbers (see Figs. 3 and 4, upper panels) and in
the magnetic field intensity (see \url{http://wso.stanford.edu/Polar.html}) 
is due to a strong decrease in
the equatorial rather than the axial dipole (in Cycle 24, the latter
even somewhat increased). This may imply that the axial and
equatorial dipoles reflect different aspects of the magnetic field
generation on the Sun.

\section{The position of the north pole of the dipole and its variations during a cycle}

Let us now trace cycle variations in the position of the north pole of the solar total dipole both in latitude and in longitude. By definition, the latitude of the pole of the axial dipole
is $\pm 90^\circ$ and the longitude is not determined. For the vector of the equatorial dipole parallel to the equatorial plane, the latitude of the pole is always zero and the longitude is a
significant characteristic. The position of the north pole of the total dipole on the sphere depends on the longitude of the horizontal dipole and the ratio of the absolute values of the
vertical and horizontal components.

Fig. 5 represents the positions of the north magnetic pole on the surface of the Sun for each Carrington rotation in Cycles 21-24 in the polar coordinates $(\varphi, \theta)$, where $\theta$ and $\varphi$ are, respectively,
the latitude and longitude of the point on the solar surface. Note that the polar regions in the figure look more stretched than the equatorial ones. The center of the coordinate system is the heliographic pole of
the corresponding hemisphere. Red color shows the position of the pole in the northern hemisphere, blue, in the southern hemisphere.

Similar diagrams for Cycles 21-23 in Livshits \& Obridko (2006) show that during the epoch of minimum, the pole of the dipole executes fairly regular precession-like motions making one or two turns relative
to the rotation pole. This quasi-precession lasts from 1 to 3 years, during which the situation of the inclined rotator is realized. After that, a sharp jump (during 0.7 - 1.2 years) occurs into the equatorial region, where
the dipole continues smooth motion along the longitude for 1.5 - 3 years. Then, a new jump occurs, and the precession continues at the opposite pole. The movement of the pole in the
equatorial zone is rather complicated. De Rosa et al. (2012) called it "aimless wandering", because it is determined by the random interaction of the active regions and the equatorial dipole.
Livshits \& Obridko (2006) revealed quasi-biennial variations in these seemingly random wanderings. Therefore, we have applied averaging over 27 Carrington rotations (corresponding quite precisely to two years) to plot
Fig.~5. With this averaging, the picture becomes much clearer; one can see full cycles described by the poles. Moreover, the transit from high to low and medium latitudes
occurs much faster and covers rather a narrow longitude range.

\begin{figure}
\includegraphics[width=0.5\columnwidth]{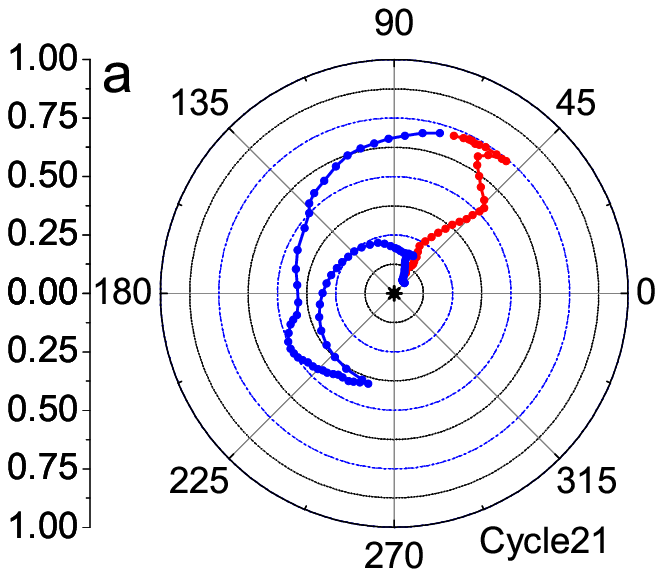}\includegraphics[width=0.5\columnwidth]{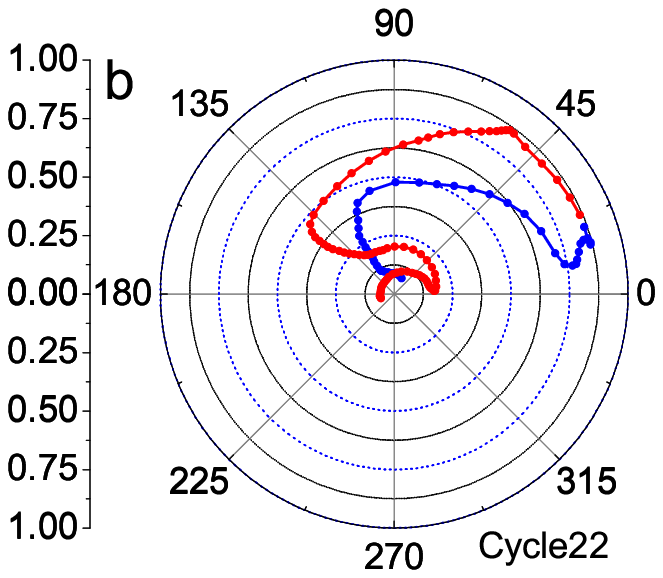}\\
\includegraphics[width=0.5\columnwidth]{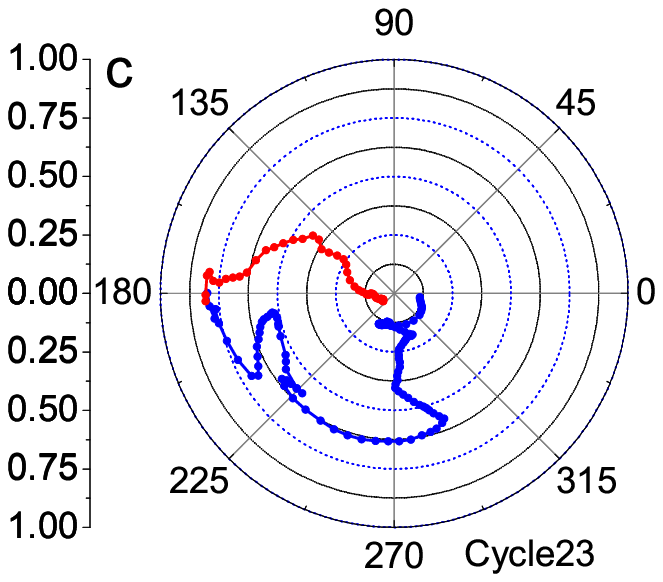}\includegraphics[width=0.5\columnwidth]{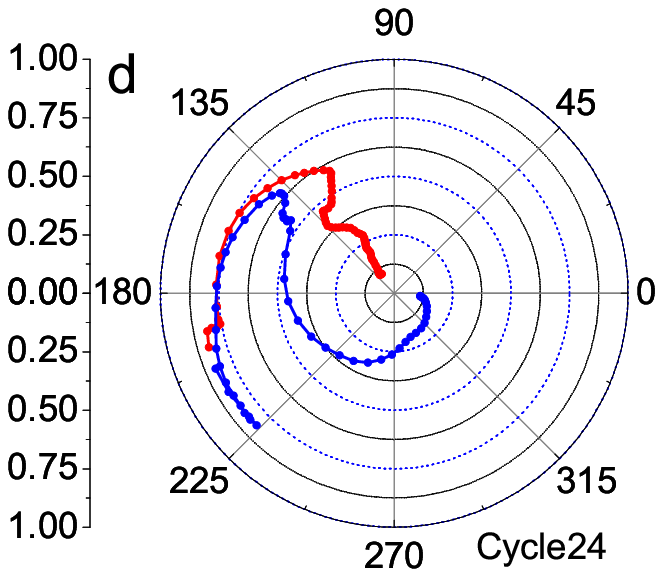}
\caption{Positions of the north pole of the magnetic dipole in four solar activity cycles. Red color corresponds to the northern hemisphere, blue, to the southern hemisphere.
The diagrams show the northern (top) and southern (bottom) hemispheres of the Sun. The circles are the contour lines of the values of $\cos \theta$ given on the vertical scales.
The values of $\varphi$ are given in degrees.  }
    \label{fig5}
\end{figure}

One can see a general counterclockwise latitudinal rotation. This means that the longitude increases, probably, due a weak acceleration of the dipole rotation relative to the Carrington coordinates.

\section{Characteristics of the quadrupole}

\begin{figure}
    \includegraphics[width=\columnwidth]{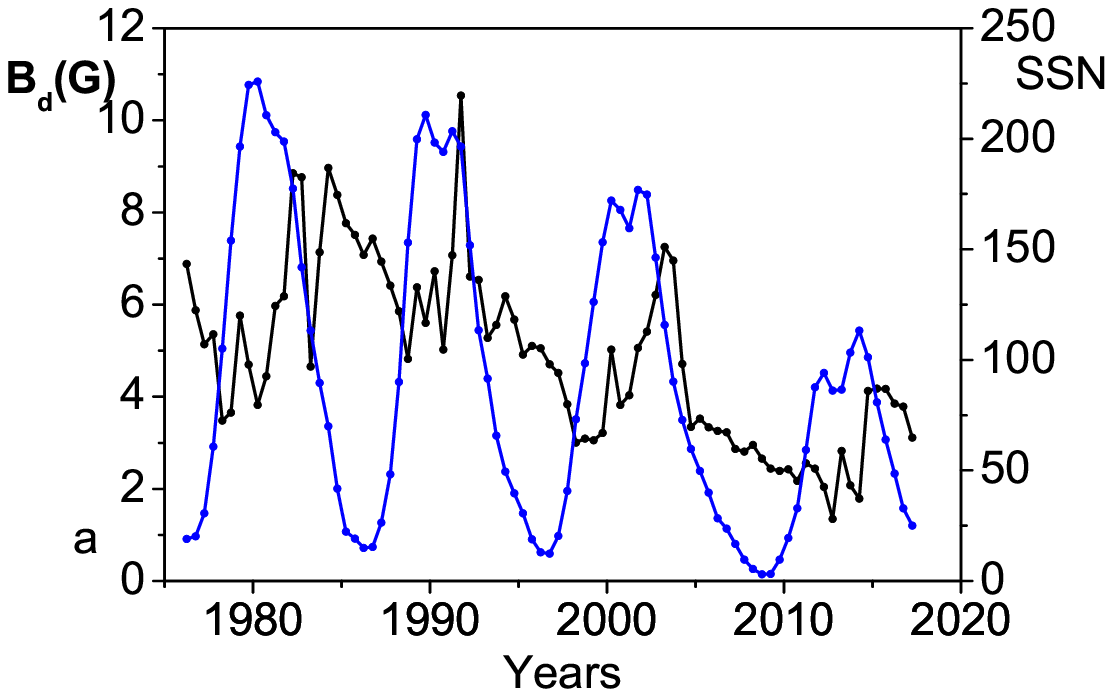}\\
    \includegraphics[width=\columnwidth]{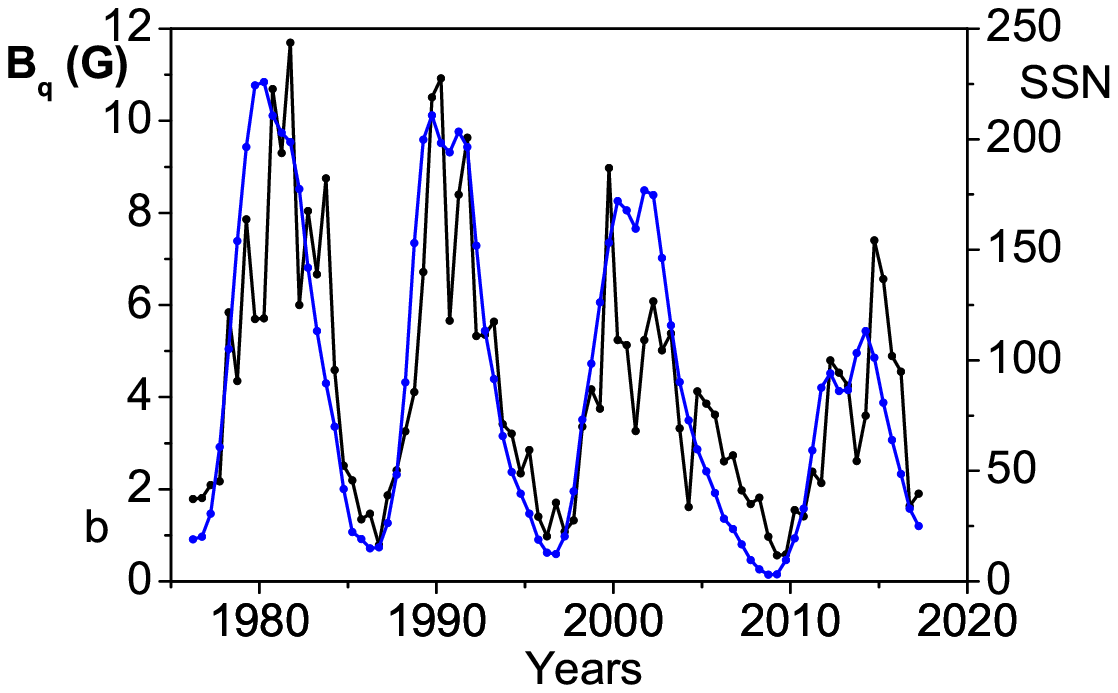}
    \caption{Mean semiannual values of the magnetic field of the solar dipole (l=1) -- black curve (a) and mean semiannual values of the magnetic pole of the solar quadrupole (l=2) -- black curve (b). The blue curves on both panels show the mean semiannual sunspot numbers (Version 2).}
    \label{fig6}
\end{figure}

Fig.~6a represents the mean semiannual unsigned values of the
magnetic field of the solar dipole ($l=1$) averaged all over the
solar surface ($B_d$). As noted above, the maximum of the mean
dipole values does not coincide in time either with the maximum of
the polar field, which is observed in the vicinity of the sunspot
minimum, or with the sunspot maximum. The point is that we
calculated the mean magnetic field of the total dipole, which
consists of both the axial and the equatorial dipoles. The vertical
(or axial dipole) peaks near the minimum of the sunspot cycle. The
horizontal (or equatorial) dipole reaches its maximum value near the
maximum of the cycle. As a result, the maximum of the total dipole
is observed at the beginning of the declining phase. Besides that,
the magnetic field of the dipole is somewhat smaller than the
maximum value of the polar field. This is due to the fact that the
intensity of the polar field is determined not only by the axial
dipole, but also by other axisymmetric (odd relative to the equator)
components.

At the sunspot maximum, the axial dipole tends to zero (the polarity reversal occurs), while the equatorial dipole still exists at low latitudes. The ratio of the maximum to minimum dipole value is approximately 3:1.

Fig.~6b shows the mean semiannual unsigned values of the magnetic
field of the solar quadrupole ($l-2$)averaged all over the solar
surface ($B_q$). The quadrupole reaches its maximum values in the
vicinity of the sunspot maximum, sometimes slightly ahead (in Cycles
22 and 23) or behind it (in Cycles 21 and 24). A dramatic difference
from the behavior of the dipole is that in the epochs of minimum,
the quadrupole field drops by an order of magnitude down to ~0.1
Gauss.

\begin{figure}
   \includegraphics[width=\columnwidth]{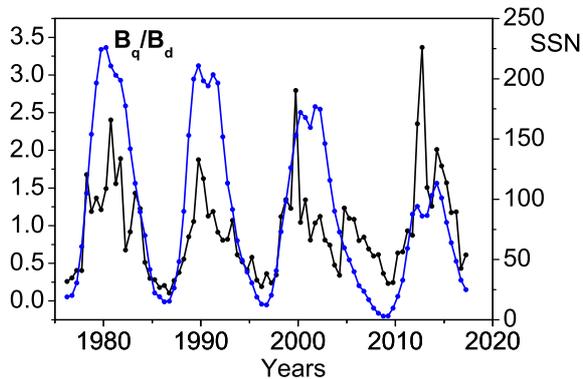}
    \caption{Ratio of the mean semiannual values of the magnetic field of the solar quadrupole $B_q$ to the mean semiannual values of the magnetic pole of the solar  dipole $B_d$. The blue curve shows sunspot numbers (Version 2).   }
    \label{fig7}
\end{figure}

During the past four cycles of activity, the maximum values of the dipole and quadrupole were decreasing. However the decrease in the quadrupole field was much smaller. As a result, the ratio $B_q/B_d$
was increasing gradually (Fig.~7) and reached ~3 in Cycle 24. This means that in Cycle 24, the quasi-symmetric dynamo generation mechanism proved to be more efficient. Note that such a relative
increase in the contribution of the total component was recorded before the Maunder minimum (Ribes\& Nesme-Ribes, 1993; Sokoloff \& Nesme-Ribes, 1994).  It looks instructive to note that
in other stars Zeeman-Doppler imaging studies have revealed that the large scale fields of stars tend to get more non-axisymmetric with increasing rotation rates (See et al., 2016)  and persistent non-axisymmetric spot structures only appear on solar-type stars with high enough activity levels (Lehtinen et al., 2016).

\begin{figure}
    \includegraphics[width=\columnwidth]{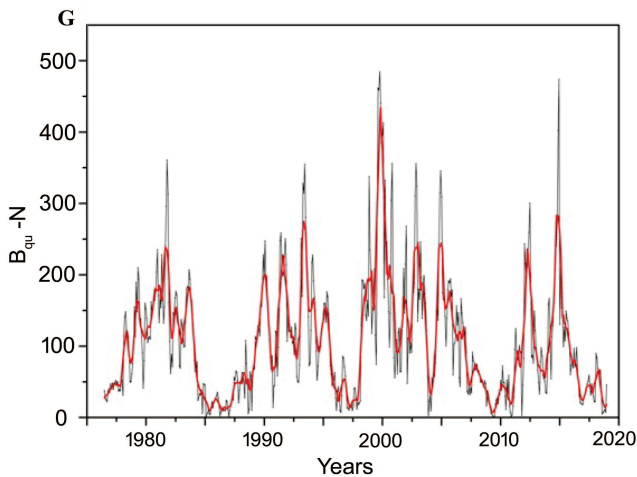}
    \caption{Magnetic field of the north pole of a quadrupole. The black curve shows these values for one of the north poles. The red curve shows the same values smoothed over seven rotations (i.e., about six months).}
    \label{fig8}
\end{figure}

Strictly speaking, the notion of the pole of a quadrupole requires a more precise definition. By analogy with the dipole, we mean by the pole of a quadrupole the point where the components $B_\varphi$ and $B_\theta$ vanish,
the field is strictly radial and reaches the maximum value. The quadrupole has four such points. In two of them the field is positive; such points will be called poles N1 and N2. Similarly,
the points with the negative ("southern") field polarity are denoted as S1 and S2.

The values of the quadrupole magnetic field at  the poles of one
sign virtually coincide. The black curve in Fig. 8 shows these
values for one of the north poles. The red curve shows the same
values smoothed over seven rotations (i.e., about six months).

One can see a fine structure with the peaks shifted relative to each
other by 2 -- 3 years. We will return to this issue in Sect.~7.

Determining the longitude of a quadrupole rotating in the Carrington coordinate system is not an easy task. This is difficult to do analytically from Eqs. (1-3). Therefore, we have used two methods.

1. One is the method of joint minimization of $B_\varphi$ and $B_\theta$. For each rotation, we computed a grid of values of all three components of the quadrupole magnetic field and, then, found the
coordinates of the four poles where these components were minimal. Unfortunately, this method may yield significant errors in the epochs of the activity minimum, when all components of the
quadrupole magnetic field are close to zero. Besides, certain difficulties arise when the longitude of a gradually moving quadrupole falls outside the range of $0-360^\circ$ and a gap in the values
occurs, which has to be somehow filled in.

2. The second method is graphical. For each rotation, we plotted a map of all components of the magnetic field (one of the maps is represented in Fig.~2b) and found preliminary coordinates visually.
Then, the coordinates were specified by calculating the field values. This method is undoubtedly more laborious, but it provides additional data control and makes it easier to eliminate the uncertainty
arising when the pole goes beyond $0-360^\circ$. Thus, we extend the notion of longitude taking into account the prehistory of excursions of the pole.

The results obtained by both methods were brought together. The fact that, for physical reasons, the longitudes of the poles $N_1$, $S_1$, $N_2$, and $S_2$ must consecutively differ by $90^\circ$, $180^\circ$ and $270^\circ$
was used for additional control.

The values of the extended longitude obtained in such a way are shown in Fig. 9. The zero-point of the longitude scale corresponds to the beginning of rotation 2025 (January 2, 2005). The total
range of changes of each pole for four rotations was about $700^\circ$, i.e., approximately two Carrington rotations.

These calculations infer two important conclusions.

1.  One can notice a gradual decrease in longitude by about $600^\circ$ in the same phases of the cycle over 42 years (i.e., over 567 rotations). This may mean a secular slowdown of the quadrupole by $1^\circ$ per rotation
(i.e. by 0.28\%). On large time scales, this effect can be significant.

2.  The rotation rate of the quadrupole depends on the phase of the cycle. On the ascending branch of the activity cycle, the longitude decreases and, therefore, the rotation rate is lower
than Carrington. After the maximum, it becomes higher than Carrington. At the minimum, the longitude does not change for some time (occasionally quite long, as it did between Cycles 22 and 23),
and the dipole takes a stable position in this coordinate system. This effect is visible in all cycles, but in Cycle 21 it is rather weak.

\begin{figure}
   \includegraphics[width=\columnwidth]{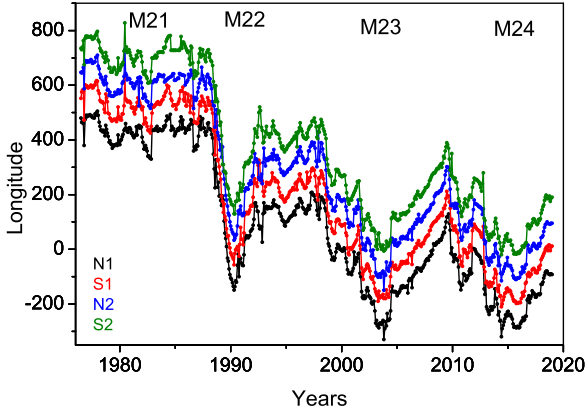}
    \caption{The longitude of four poles of a quadrupole.}
    \label{fig9}
\end{figure}

This may be due to the fact that the pole of a quadrupole moves in latitude. Fig.~10 shows the relationship between the longitude and latitude of the pole. To simplify the picture, we smoothed the data over
27 rotations, i.e., over a window of about two years. In doing so we were taking into account the absolute value of the latitude, which is indicated in the figure in degrees. The generalized longitude
is reduced to the range $0-360^\circ$. One can see that at the beginning of the cycle, the pole moves clockwise (the rotation rate decreases); in the second part of the cycle, the direction of motion is reversed,
i.e., the rotation rate increases. Moreover, the pole in the first part of the cycle is mostly located at high latitudes, where the rotation rate calculated following the differential rotation law is lower
than the Carrington velocity, while in the second part, the latitude drops. It is difficult to say whether the variations in the quadrupole rotation rate are limited to this effect, since the quadrupole
itself is not a direct result of the existence of active regions.

\begin{figure}
    \includegraphics[width=\columnwidth]{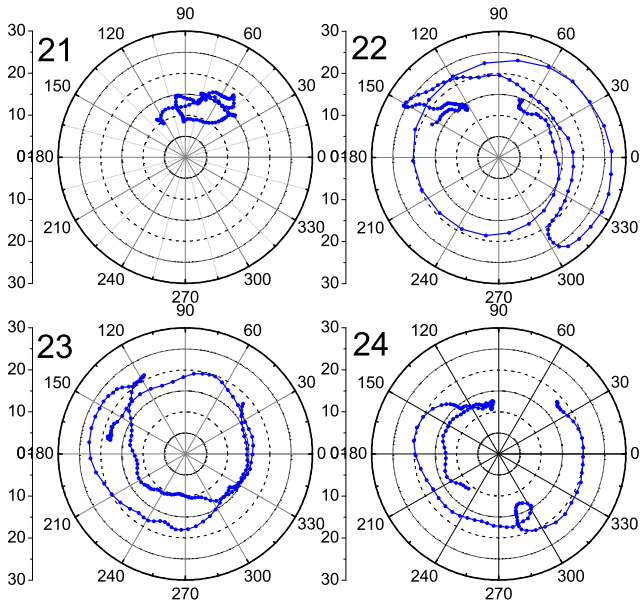}
    \caption{Relationship between the longitude and latitude of the pole of a quadrupole.}
    \label{fig10}
\end{figure}

\begin{figure*}
    \includegraphics[width=\textwidth]{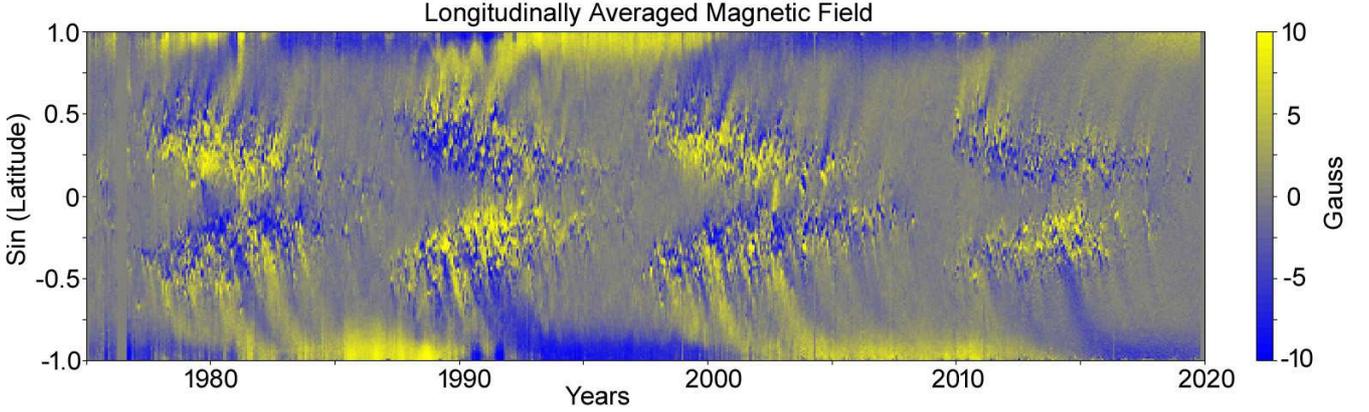}
    \caption{Distribution of the surface magnetic field (longitudinally averaged) over the last four solar cycles (from
\url{http://solarcyclescience.com/solarcycle.html}).}
    \label{fig11}
\end{figure*}

The Carrington coordinates we are using in the Sun are referred
to the latitude of 16$^{\circ}$. For latitudes closer to the equator,
the rotation period is smaller (i.e., higher rotation rate), while
for higher latitudes, the effect is {\it vice versa}. Fig. 11
represents the distribution of the longitudinally averaged surface
magnetic field over the last four solar cycles (from
\url{http://solarcyclescience.com/solarcycle.html}). We have added two
solid black lines to the Figure to show the Carrington latitudes.
The shift of the center of gravity of the activity to the equatorial
zone results in a decrease of the period and the corresponding
longitude in Figs.\ 9 and 12. However, this effect cannot fully
explain differences in the rotation of the dipole and quadrupole.

\section{Rotation of the dipole and quadrupole}

As mentioned above, a secular deceleration of the quadrupole rotation is observed over a long time interval. In the case of the dipole, this effect is absent or has the opposite sign.

\begin{figure}
    \includegraphics[width=\columnwidth]{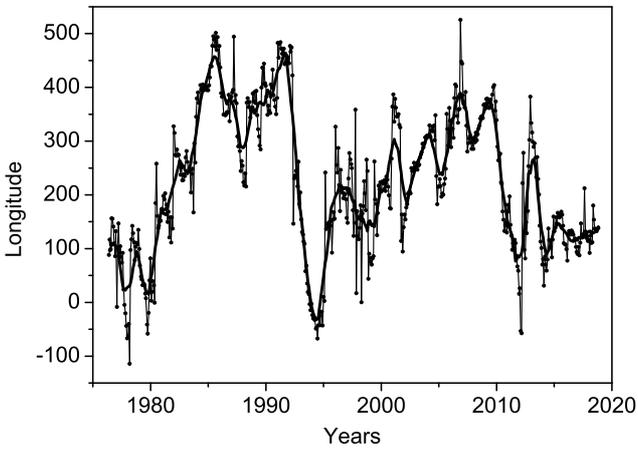}
    \caption{Longitude of the dipole north pole.}
    \label{fig12}
\end{figure}

For 16 years from 1976 to 1992, the longitude of the dipole pole was gradually increasing; i.e., the rotation rate exceeded the Carrington velocity by about 0.6\%. After that, a very fast
reconstruction occurred followed by a return to the previous position and, again, the accelerated rotation at about the same velocity.

Against this background, both the dipole and the quadrupole display minor velocity fluctuations. The position of the longitude as a function of time is shown in Fig.~12.

\begin{figure}
   \includegraphics[width=0.95\columnwidth]{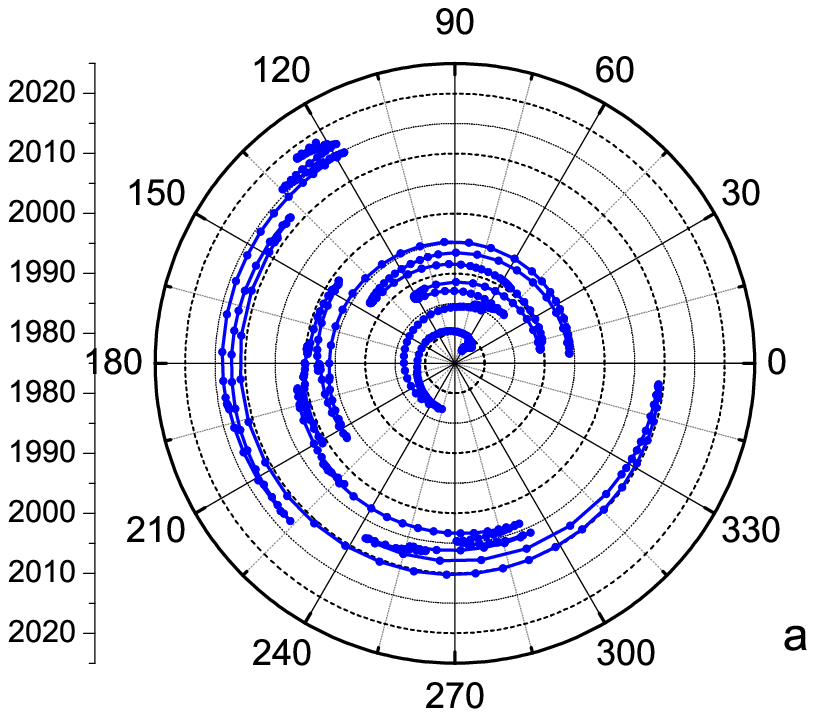}
   \includegraphics[width=\columnwidth]{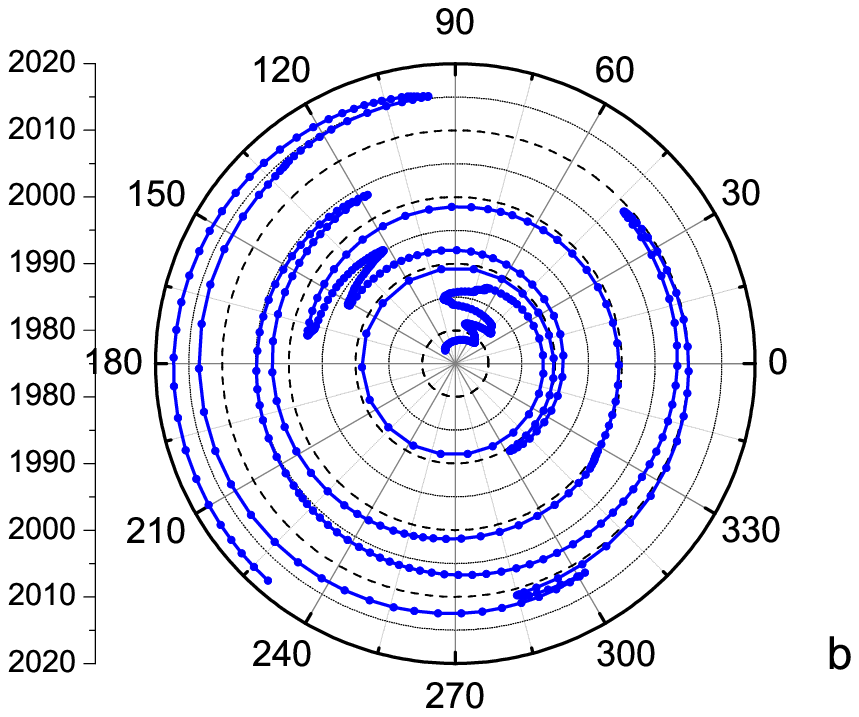}
   \caption{Longitude-time diagram for the dipole (a) and quadrupole (b).}
   \label{fig13}
\end{figure}

The diagrams in Fig.\ 13 seem very much alike, but upon careful
examination together with Figs.\ 9 and 12 it becomes clear that, the
dipole rotates predominantly counterclockwise, while the quadrupole,
changes the direction of the rotation many times. This reflects the
rotation of these two multipoles in the Carrington system. That is,
in addition to the main rotation modes, there is also a finer
structure. Variations in the rotation rate can be studied directly
from the data represented in Figs. 9 and 12.

Let us introduce the parameter

\begin{equation}
R(i) = (\varphi(i+1) - \varphi(i-1))/720,
\label{eq4}
\end{equation}
where $i$ is the number of the following Carrington rotation and $\varphi(i)$ is the longitude. The parameter $R$ introduced in this equation means the percentage deviation from the Carrington rotation rate over a small time interval. Its positive value corresponds to a higher rotation rate and negative, to a lower rotation rate.

The obtained values smoothed over 14 rotations are shown in Fig.~14.

\begin{figure}
    \includegraphics[width=\columnwidth]{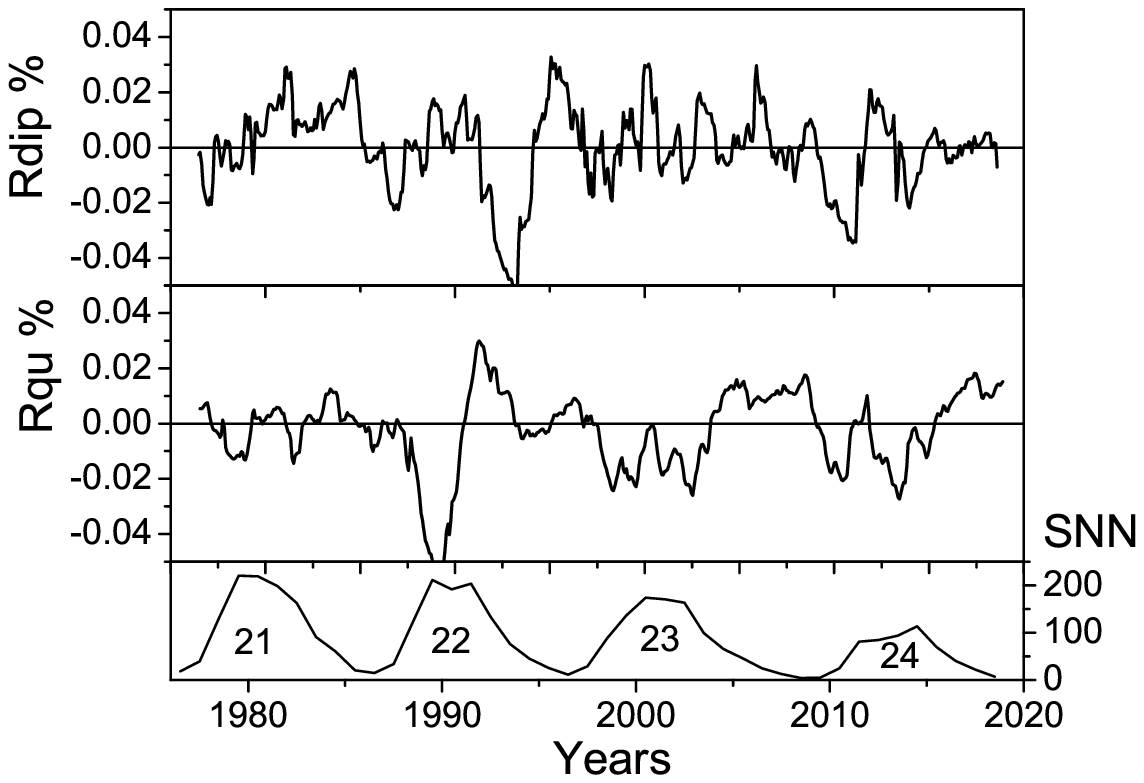}
    \caption{Variations in the dipole and quadrupole rotation rates.}
    \label{fig14}
\end{figure}

Fig. 14 shows that variations in the rotation rate of a quadrupole
range from 0,02\% to 0.06\% and display a relationship with the
activity cycle. Parameter R is usually negative at the beginning and
at the maximum of the cycle and becomes positive in the declining
phase and at the minimum. In the case of a dipole, the variations in
the same range do not show clear connection with the cycle being
rather chaotic.

Thus, in addition to the secular deceleration of the quadrupole by 0.28\% per rotation, we can also observe cyclic variations of five to ten times smaller scales.
In the case of the dipole (see above), there are long periods when the rotation rate exceeds the Carrington velocity by about 0.6\% and
chaotic variations by an order of magnitude weaker are observed.

Comparing our findings with available stellar data, we note that
\cite{Cetal14} found secular azimuthal dynamo waves in the dipolar field component in their numerical simulations which show some resemblance to the secular trend seen in the solar quadrupolar longitudes. The retrograde migration trend of the solar quadrupole is also reminiscent of the similar migration trends found by \cite{Letal16}  for the stellar active longitudes, although they reported generally prograde migration. Of course, a self-consistent comparison of stellar and solar data has to de addressed in a particular paper.

\section{Discussion and Conclusions}

In this work, we have considered variations of the main components of the large-scale magnetic field: the dipole (axial and equatorial, separately) and the quadrupole.
It is shown that the three main components of the large-scale field behave differently and are not manifestations of the same process.

The axial dipole cannot be fully identified with the polar field; its moment is much lower than the value characteristic of the polar field. The magnitude of the polar field
depends not only on the axial dipole, but also on higher-order odd harmonics. At the polarity reversal, the axial dipole, naturally, vanishes. However, the period of the polarity
reversal does not coincide with that of the cycle maximum, as is commonly believed, but occurs 0.5-1 years earlier and takes a very short time.

The characteristics of the equatorial dipole depend to a certain extent (but not entirely) on the contribution of magnetic fields of the solar active regions. This contribution is, naturally,
reduced, since, according to the Hale laws, the fields of active regions are associated with the toroidal field, which is even with respect to the equator. The maximum of the equatorial dipole
is shifted from the maximum of the cycle towards the beginning of the declining phase and is quite clearly pronounced. Both the axial and the equatorial dipoles have exhibited a systematic decrease during the past four cycles in agreement with the general decrease of solar activity.

The transition of the pole of a dipole from the polar region to mid latitudes occurs rather quickly, so that the longitude of the pole changes little.
With time, however, this inclined dipole region shifts to higher longitudes, which suggests a weak acceleration of the dipole rotation. The mean rotation rate exceeds the
Carrington velocity by 0.6\%. Perhaps this is due to the fact that the dipole rotation in longitude is determined strictly by the equatorial dipole. According to
the differential rotation law, the Carrington velocity corresponds to a latitude of $\pm 26^\circ$, while closer to the equator, the rotation rate increases. At shorter
time intervals, the dipole displays a random wandering not related in any way to the phase of the cycle.

The behavior of the quadrupole differs dramatically. Although the quadrupole moment has also decreased for the past four cycles, this decrease was much smaller than that of
the dipole moment. As a result, the ratio of the quadrupole and dipole moments has increased for four cycles more than twice in contrast to the sunspot numbers, which
displayed a twofold decrease for the same time interval. This reminds us of the situation before the Maunder grand minimum, which, to judge from some indices, had been preceded by an enhancement of the even component of the magnetic field.

It should be noted that using the values smoothed over 7 rotations (approximately half a year), as in Fig.~8, instead of the mean semiannual values, as in Fig.~6, makes the picture even more interesting.
The maxima of the cycles appear as a series of peaks separated by profound dips. The height of the peaks does not decrease at all with time.

The rotation pattern of the quadrupole is more complex and, at the same time, more orderly.

In general, the mean longitude of the poles of the same sign
decreased by 600 degrees over four cycles, which implies that the
mean rotation rate was lower than the Carrington velocity by 0.28\%.
The secular deceleration is imposed by cyclic variations with the
rotation rate being somewhat lower in the rise phase (by $\approx
0.06$\%) and becoming positive in the decline phase. These cyclic
variations may also be due to the fact that, before the maximum, the
pole of the quadrupole is located at mid-latitudes, where the
rotation rate is lower than the Carrington velocity, and after the
maximum, it is mainly located near the equator, where the rotation
rate is higher than the Carrington velocity. The secular
deceleration may be the result of a gradual shift of the
high-latitude boundary between the old and new fields towards lower
latitudes, where the rotation rate is lower. For the past four
cycles, this boundary moved from $30^\circ$ to $20^\circ$
\url{https://solarscience.msfc.nasa.gov/images/magbfly.jpg}. Now, let us
see what our analysis says about the symmetry of the magnetic fields
generated by the solar dynamo.

On the whole,  solar quadrupole behaves quite differently from the solar dipole and contributes substantially into the total solar magnetic field, however we do not see specific periods associated with solar quadrupole at least
in the times under discussion. Correspondingly, it looks plausible that a mode of quadrupole symmetry was not excited on the Sun along with the dipole mode at least in the times under discussion.
Perhaps the most convincing evidence is that the value of the quadrupole moment varies with the same 11-year period as the dipole moment. It seems reasonable to suggest that other differences in the behavior of the dipole and quadrupole moments are due to the fact that deviations of the solar hydrodynamics from strict symmetry about the equator have different effect on the dipole and quadrupole components.
In any case, we do not see here such dramatic phenomena as the occurrence of sunspots almost exclusively
in one solar hemisphere, as was observed at the end of the Maunder minimum.

Finding out how minor deviations from the symmetry about the equator lead to differences in the behavior of the dipole and quadrupole fields and how frequent these deviations
are in the Sun would require special analysis of observations and numerical modeling, which go far beyond the scope of this article.

At the same time, we cannot ignore the fact that the current behavior of the solar magnetic field differs noticeably from what was observed over the past century (e.g.  de Jager et al., 2016).
So it is possible that the Sun is approaching a new Grand Minimum. We have no opportunity to check whether the particularities of the quadrupole behavior described above were typical throughout
the past century. On the other hand, we do not know what exactly happened on the Sun just before the Maunder minimum, since the observation data available are not very definite (see
Usoskin et al., 2015; Zolotova \& Ponyavin; 2015). However, these data also sometimes display signatures of significant asymmetry in the distribution of sunspots relative to the solar equator
(Nesme-Ribes et al., 1994).

Notwithstanding all limitations, we can conclude that at present there is no reason to believe that either quadrupole or mixed parity magnetic configurations are excited in the Sun along with
the main dipole mode. The occurrence of the quadrupole and other even harmonics is more rightfully attributed to deviations of the solar hydrodynamics from a strict symmetry with respect to
the solar equator.

\section*{Acknowledgements}

We acknowledge the help of the Wilcox Solar Observatory. Data used
in this study was obtained via the web site \url{http://wso.stanford.edu/},
courtesy of J.H.Hoeksema. The work was supported by the RFBR
grants 17-02-00300, 19-02-00191a, 19-52-53045 and 18-02-00085.







\bsp    
\label{lastpage}
\end{document}